\documentclass{article}
\usepackage{amsmath,amsthm}
\usepackage{amssymb,latexsym}
\usepackage[mathscr]{eucal}
\usepackage{setspace}
\usepackage{graphics}
\usepackage{array}

\newcommand{\lp}{\left(}
\newcommand{\rp}{\right)}
\newcommand{\lb}{\left[}
\newcommand{\rb}{\right]}
\newcommand{\lc}{\left\{}
\newcommand{\rc}{\right\}}

\newcommand{\be}{\begin{equation}}
\newcommand{\ee}{\end{equation}}

\begin{document}

\title{\textbf{Parker's Stellar Wind Model for Polytropic Gas Flows}}         
\author{B. K. Shivamoggi and D. K. Rollins\\
University of Central Florida\\
Orlando, FL 32816-1364, USA\\
}        
\date{}          
\maketitle

\noindent \large{\bf Abstract} \\ \\
Parker's \cite{par} hydrodynamic stellar wind model is extended to polytropic gas flows. A compatible theoretical formulation is given and detailed numerical and systematic asymptotic theoretical considerations are presented. The polytropic conditions are shown to lead to tenuous and faster wind flows and hence enable the stars to lose their angular momentum more quickly.

\pagebreak

\noindent\Large\textbf{1. Introduction}\\

\large Stellar wind is a continuous plasma outflow from a star (in the case of the Sun, this outflow\footnote{SOHO observations (Cho et al. \cite{cho}) revealed that the structure of the solar wind changes from the solar maximum to the solar minimum period.} typically emerges from coronal holes (Sakao et al. \cite{sak}) and carries a remnant of the stellar magnetic field that fills the space around the star (in the case of the Sun, this constitutes the heliosphere (Dialynas et al. \cite{dia})). Stellar winds carry off, especially when magnetized, a huge amount of angular momentum from the stars while causing a very negligible amount of mass loss from the stars. Weak to moderate stellar winds are generated by an expanding outer corona due to an extended active heating of the corona\footnote{Parker \cite{par}-\cite{par2} proposed that this is caused by the dissipation of plasma waves produced by microflares in the coronal holes. The details of the coronal heating mechanism are still controversial (Klimchuk \cite{kli}).} in conjunction with high thermal conduction. However, control of coronal base conditions and high thermal conduction are both inadequate to generate observed high wind speeds in the case of the Sun (Parker \cite{par4},\cite{par5}). This indicates the rationale for some additional acceleration mechanism to operate beyond the coronal base. Parker \cite{par4} gave an ingenius stationary model which provided for the smooth acceleration of the stellar wind through transonic speeds by continually converting the thermal energy into the kinetic energy of the wind. In the case of the Sun, the solar wind was confirmed and its properties were recorded by {\it in situ} observations (Neugebauer and Snyder \cite{neu}, Hundhausen \cite{hun}, Meyer-Vernet \cite{mey}). The stellar rotation is found to lead to faster stellar winds and hence enable protostars and strong rotators to lose their angular momentum quickly via the mechanism of centrifugal and magnetic driving (Shivamoggi \cite{shivamoggi}).

One of the main assumptions in Parker's \cite{par4} model is that the gas flow occurs under isothermal conditions (in standard rotation), 

\be
p=a_0^2\rho
\ee

\noindent where $a_0$ is the constant speed of sound. However, the extended active heating of the corona may be represented in a first approximation by using the polytropic gas \footnote{ Polytropic model provides a more general framework and is fully compatible with the adiabatic condition which is a special case of the polytropic model.} relation (Parker \cite{par6}, Holzer \cite{hol}, Keppens and Goedbloed \cite{kep}),

\be
p=C\rho^\gamma
\ee

\noindent where $C$ is an arbitrary constant and $\gamma$ is the polytropic exponent, $1<\gamma<5/3$. The modified Parker's equation governing the acceleration of stellar wind of a polytropic gas given by Holzer \cite{hol} seems to be erroneous. The purpose of this paper is to rectify this error and present a more compatible formulation and then make detailed numerical and systematic asymptotic theoretical considerations to describe acceleration of stellar wind of a polytropic gas. The polytropic gas conditions appear to lead to tenuous and faster wind flows and hence enable the stars to lose their angular momentum more quickly.

\newpage

\vspace{0.3in}

\noindent\Large\textbf{2. Polytropic Gas Stellar Wind Model}\\

\large In Parker's hydrodynamic model \cite{par4} the stellar wind is represented by a steady and spherically symmetric\footnote{In reality, observations of the solar wind (Wang and Sheeley \cite{wan}) suggested and (Kopp and Holzer \cite{kop}) proposed a rapidly-diverging superradial wind flow especially in some active regions like the coronal holes.} flow so the flow variables depend only on $r$, the distance from the star. The flow velocity is further taken to be only in the radial direction - either inward (accretion model) or outward (wind model). We assume for analytical simplicity that the flow variables and their derivatives vary continuously so there are no shocks anywhere in the region under consideration.

The mass conservation equation is

\be
\frac{2}{r}+\frac{1}{\rho}\phantom{x}\frac{d\rho}{dp}+\frac{1}{v_r}\phantom{x}\frac{dv_r}{dr}=0.
\ee

Assuming the gravitational field to be produced by a central mass $M_s$, Euler's equation of momentum balance is 

\be
\rho v_r\frac{dv_r}{dr}=-~\frac{dp}{dr}-\frac{GM_s}{r^2}\rho
\ee

\noindent $G$ being the gravitational constant.

Using equations (2) and (3), equation (4) becomes

\be
\lp M^2-1\rp\frac{2}{v_r}\phantom{x}\frac{dv_r}{dr}=\frac{4}{r^2}\lp r-r_\ast\rp
\ee

\noindent where,

\be
a^2\equiv\frac{dp}{d\rho}=\frac{\gamma p}{\rho}, M\equiv\frac{v_r}{a}, r_\ast\equiv\frac{GM_s}{2a^2} 
\ee

Using (6), and noting the relation, \footnote{(\ref{eq::7}) is universally valid because it simply represents the energy conservation condition.}

\be
\frac{a^2}{a^2_0}=\frac{1}{1+\lp\frac{\gamma-1}{2}\rp M^2} 
\label{eq::7}
\ee

\noindent equation (5) becomes, 

\be
\frac{\lp M^2-1\rp}{M^2\lb 1+\lp\frac{\gamma-1}{2}\rp M^2\rb}\phantom{x}\frac{d}{dr}\lp M^2\rp=\frac{4}{r^2}\lp r-r_\ast\rp.
\ee

\vspace{.10in}

By introducing the total energy,

\be
E\equiv\frac{v_r^2}{2}+\frac{a^2}{\gamma-1}-\displaystyle\frac{GM_s}{r}
\ee

\noindent The right hand side of equation (8) may be written alternatively as

\be
\frac{\lp M^2-1\rp}{M^2\lb 1+\lp\frac{\gamma-1}{2}\rp M^2\rb}\phantom{x}\frac{d}{dr}\lp M^2\rp=\frac{4E+\lp\frac{4\gamma-6}{\gamma-1}-M^2\rp\displaystyle\frac{GM_s}{r}}{r\lp E+\displaystyle\frac{GM_s}{r}\rp}
\ee

\noindent which shows that the corresponding result given by Holzer \cite{hol} is apparently erroneous. 

Introducing, 

\be
r_{\ast_0}\equiv\displaystyle\frac{GM_s}{a^2_0}
\ee
 
\noindent and using (7), equation (8) may be rewritten as

\be
\frac{\lp M^2-1\rp}{M^2\lb 1+\lp\frac{\gamma-1}{2}\rp M^2\rb}\phantom{x}\frac{d}{dr}\lp M^2\rp=\frac{4}{r^2}\lb r-\lc 1+\lp\frac{\gamma-1}{2}\rp M^2\rc r_{\ast_0}\rb.        
\label{eq::12}
\ee
The numerical solution of equation (\ref{eq::12}) is plotted in figure 1, which shows the polytropic conditions enhance the acceleration of stellar winds.

We now consider some special asymptotic cases for which equation (12) facilitates simpler solutions.

\vspace{0.3in}

\noindent\Large\textbf{(i) Transonic Regime}\\

\large Near the sonic critical point $r=r_\ast$, we may write,

\be
r=r_\ast+x,~M^2=1+y.
\ee

Using (7) and (11), equation (12) then gives 

\be
y\frac{dy}{dx}\approx\frac{8}{r^2_{\ast_0}\lp\gamma +1\rp}x
\ee

\noindent from which,

\be\tag{15a}
y\approx\frac{2/r_{\ast_0}}{\sqrt{\lp\gamma +1\rp/2}}x
\ee

\noindent or

\be\tag{15b}
\lp M^2-1\rp\approx\sqrt{2\lp\gamma +1\rp} \lp\frac{r}{r_\ast}-1\rp.
\ee

\noindent (15b) shows an enhanced acceleration of the polytropic wind $\lp \gamma >1\rp$ past the sonic critical point, in agreement with the numerical solution of equation (\ref{eq::12}).

\vspace{.10in}

\noindent\Large\textbf{(ii) Near-star Regime}\\

\large For $r/r_\ast <<1$, equation (12) may be approximated by

\be\tag{16}
\frac{1}{M^2\lb 1+\lp\frac{\gamma -1}{2}\rp M^2\rb^2}\phantom{x}\frac{d}{dr}\lp M^2\rp\approx\displaystyle\frac{4r_{\ast_0}}{r^2}
\ee

\noindent from which,

\be\tag{17}
\frac{M^2}{1+\lp\frac{\gamma -1}{2}\rp M^2}\phantom{x} \approx e^{-\displaystyle\lp 4r_{\ast_0}/r\rp}.
\ee

\vspace{.40in}

In the isothermal limit $\lp\gamma\rightarrow 1\rp$, (17) leads to 

\be\tag{18}
M^2\approx e^{-\displaystyle\lp 4r_\ast/r\rp}.
\ee

\noindent Comparison of (17) with (18) shows again an enhanced acceleration of the polytropic wind $\lp\gamma >1\rp$ near the star.

\vspace{.10in}

\noindent\Large\textbf{(iii) Far-star Regime}\\

\large For $r/r_\ast >>1$, equation (12) may be approximated by

\be\tag{19}
\frac{d}{dr} \lp M^2\rp\approx\frac{2\lp\gamma -1\rp}{r}M^2 
\ee

\noindent from which,

\be\tag{20}
M^2\approx r^{\lb 2\lp\gamma -1\rp\rb}.
\ee

\noindent (20) shows again an enhanced acceleration of the polytropic wind $\lp\gamma >1\rp$ far from the star.

\vspace{.10in}

\noindent\Large\textbf{(iv) Effective de Laval Nozzle for the Polytropic Stellar Wind}\\

\large The continuous acceleration of the polytropic stellar wind flow, seen above, from subsonic speeds at the coronal base to supersonic speeds away from the star implies a {\it de Laval} nozzle type situation operational near the star (Clauser \cite{clauser}).

If $A=A\lp r\rp$ is the cross section area of an effective de Laval nozzle associated with the polytropic stellar wind flow, we have from equations (5) and (8), 

\be\tag{21}
\begin{matrix}
\displaystyle\frac{\lp M^2-1\rp}{M^2}\lp\displaystyle\frac{2v_r}{a^2}\rp\displaystyle\frac{dv_r}{dr}=\displaystyle\frac{\lp M^2-1\rp}{M^2\lb 1+\lp\displaystyle\frac{\gamma -1}{2}\rp M^2\rb}\phantom{x}\displaystyle\frac{d}{dr}\lp M^2\rp\\
\\
=\displaystyle\frac{2}{A}\phantom{x}\displaystyle\frac{dA}{dr}=\displaystyle\frac{4}{r^2}\lp r-r_\ast\rp 
\end{matrix}
\ee

\noindent from which, 

\be\tag{22}
 A\lp r\rp=4\pi r^2e^{-2r_{\ast_0}\int^r_{r_o}\frac{1}{r^2}\lb 1+\lp\frac{\gamma-1}{2}\rp M^2\rb dr}
\ee

\noindent where $r=r_0$ at the coronal base. (22) implies that the cross section area of the effective de Laval nozzle, for the polytropic wind $\lp\gamma >1\rp$, varies faster, in consistency with the enhanced acceleration of the polytropic stellar wind.

\vspace{.30in}

\noindent\Large\textbf{5. Discussion}\\

\large Thanks to the extended heating of the corona, a polytropic gas model is in order in dealing with stellar winds. In recognition of this, the present paper is aimed at putting forward a compatible theoretical formulation and providing a detailed numerical and systematic asymptotic theoretical considerations to describe acceleration of stellar wind of a polytropic gas. The polytropic conditions have been shown to lead to tenuous and faster wind flows and hence enable the stars to lose this angular momentum more quickly.

\section*{Acknowledgments}
My thanks to Zoe Barbeau for her help with Figure 1.

\begin{figure}
	\centering
	\includegraphics{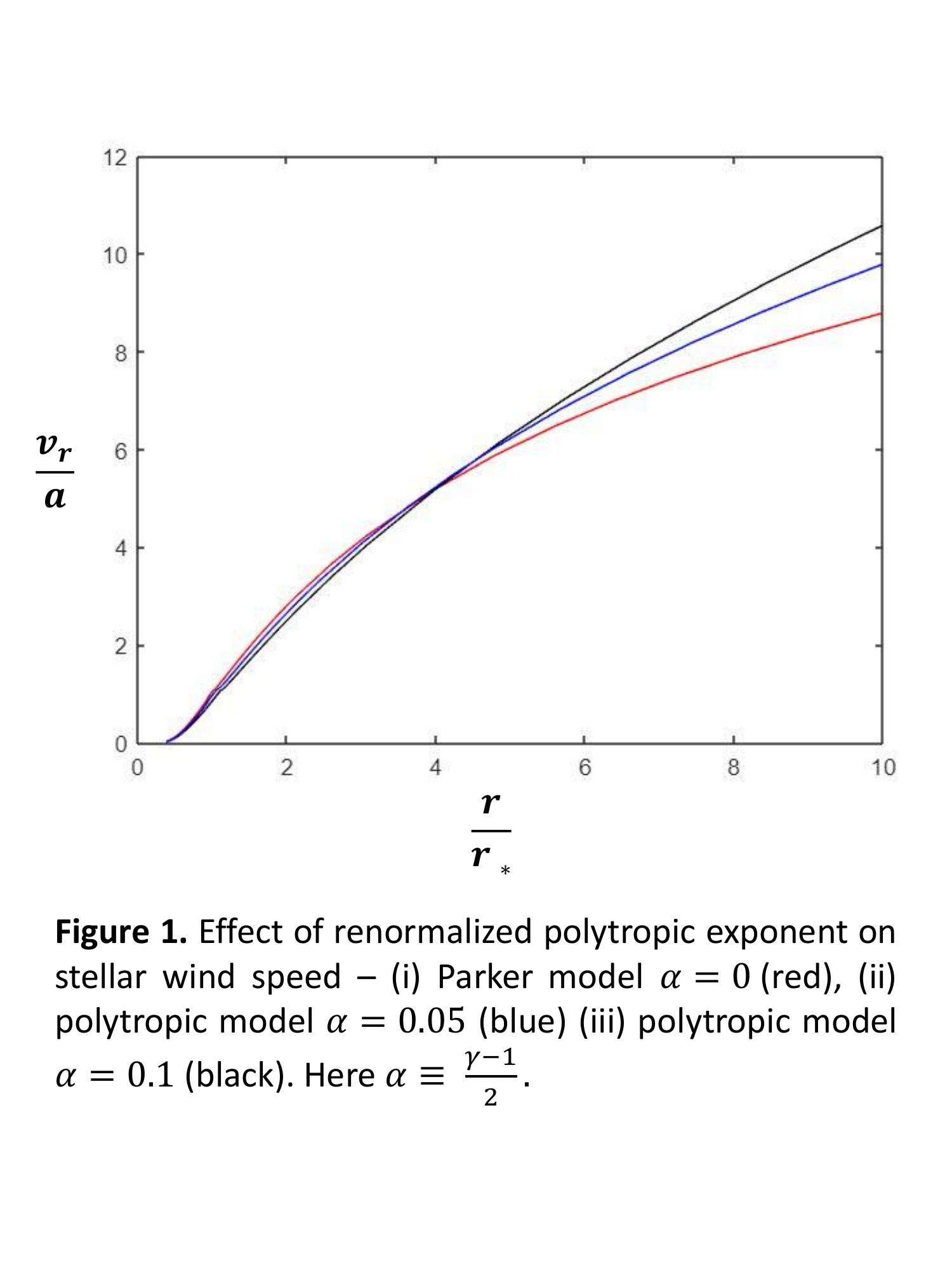}
	\label{fig:my_label}
\end{figure}


\begin{thebibliography}{xxx}

\bibitem{cho} B. H. Cho, Y. J. Moon, V. M. Nakariakov, S. C. Bong, J. Y. Lee, D. Song, H. Lee and K. S. Cho: {\it Phys. Rev. Lett.} {\bf 121}, 075101, (2018).
\bibitem{sak} T. Sakao, R. Kano, N. Narukage, J. Kotoku, T. Bando, E. DeLuca, L. Lundquist, S. Tsuneta, L. Harra, Y. Katsukawa, M. Kubo, H. Hara, K. Matsuzaki, M. Shimijo, J. A. Bookbinder, L. Golub, K. E. Korreck, Y. Su, K. Shibasaki, T. Shimizu and I. Nakatani: {\it Science} {\bf 318}, 1585, (2007).
\bibitem{dia} K. Dialynas, S. M. Krimigis, D. G. Mitchell, R. B. Decker and E. C. Roelof: {\it Nature Astron} {\bf 1}, 0115, (2017).
\bibitem{par} E. N. Parker: {\it Astrophys. J.} {\bf 264}, 642, (1983).
\bibitem{par1} E. N. Parker: {\it Astrophys. J.} {\bf 330}, 471, (1988).
\bibitem{par2} E. N. Parker: {\it Astrophys. J.} {\bf 318}, 876, (1987).
\bibitem{kli} J. A. Klimchuk: in {\it Solar Phys.} {\bf 234}, 41, (2006).
\bibitem{par4} E. N. Parker: {\it Astrophys. J.} {\bf 128}, 664, (1958).
\bibitem{par5} E. N. Parker: {\it Space Sci. Rev.} {\bf 4}, 666, (1965).
\bibitem{neu} M. Neugebauer and C. W. Snyder: {\it J. Geophys. Res.} {\bf 71}, 4469, (1966).
\bibitem{hun} A. J. Hundhausen: {\it Coronal Expansion and Solar Wind} Cambridge University Press, (1972).
\bibitem{mey} N. Meyer-Vernet: {\it Basics of the Solar Wind}, Cambridge University Press, (2007).
\bibitem{shivamoggi} B. K. Shivamoggi:  \textit{Phys. Plasmas} \textbf{27}, 012902, (2020).
\bibitem{par6} E. N. Parker: {\it Rev. Geophys. Space Phys.} {\bf 9}, 825, (1971).
\bibitem{hol} T. E. Holzer: In {\it Solar System Plasma Phys.}, Vol. 1 (ed. C. F. Kennel, L. J. Lanzerotti and E. N. Parker), North-Holland, (1979).
\bibitem{kep} R. Keppens and J. P. Goedbloed: {\it Astron. Astrophys.} {\bf 343}, 251, (1999).
\bibitem{wan} Y. M. Wang and N. R. Sheeley, Jr.: {\it Astrophys. J.} {\bf 355}, 726, (1990).
\bibitem{kop} R. A. Kopp and T. E. Holzer: {\it Solar Phys.}, {\bf 49}, 43 (1976).
\bibitem{clauser} F. H. Clauser: in \textit{4th Symp. Cosmical Gas Dynamics}, North-Holland, (1960).

\end{thebibliography}
\end{document}